\input{aipcheck}

\documentclass[
    ,final            % use final for the camera ready runs
  ]
  {aipproc}

\layoutstyle{6x9}
\begin{document}
\def\gsim{ \lower .75ex \hbox{$\sim$} \llap{\raise .27ex \hbox{$>$}} }
\def\lsim{ \lower .75ex\hbox{$\sim$} \llap{\raise .27ex \hbox{$<$}} }
\def\crexp{{\rm\thinspace km^{2} \thinspace sr \thinspace yr}}
\def\ergcms{{\rm\thinspace erg \thinspace cm^{-2} \thinspace s^{-1}}}
\def\kmps{{\rm\thinspace km \thinspace s^{-1}}}
\def\mpc{{\rm\thinspace Mpc}}
\def\ev{{\rm\thinspace eV}}
\def\kev{{\rm\thinspace keV}}
\def\hi{{\rm H}\,{\small\rm I}}
\def\sc{Schwarzschild}

\title{Blazars in hard X--rays}


\author{Gabriele Ghisellini}{
  address={Osservatorio Astronomico di Brera, Via Bianchi 46, I--23807 Merate, Italy}
}

%95.30.Gv 	Radiation mechanisms; polarization
% 95.55.Ka 	X- and γ-ray telescopes and instrumentation
%95.85.Nv 	X-ray
%95.85.Pw 	γ-ray
%98.54.Cm 	Active and peculiar galaxies and related systems 
%                 (including BL Lacertae objects, blazars, Seyfert galaxies, 
%                  Markarian galaxies, and active galactic nuclei)
%98.62.Nx 	Jets and bursts; galactic winds and fountains

\begin{abstract}
Although blazars are thought to emit most of their 
luminosity in the $\gamma$--ray band, there are
subclasses of them very prominent in hard X--rays.
These are the best candidates to be studied by Simbol--X.
They are at the extremes of the blazar sequence,
having very small or very high jet powers.
The former are the class of TeV emitting BL Lacs, 
whose synchrotron emission often peaks at tens of keV or more. 
The latter are the blazars with the most powerful jets,
have high black hole masses accreting at high 
(i.e. close to Eddington) rates.
These sources are predicted to have their high energy
peak even below the MeV band, and therefore are 
very promising candidates to be studied with Simbol--X.
\end{abstract}

\maketitle

%%%%%%%%%%%%%%%%%%%%%%%%%%%%%%%%%%%%%%%%%%%%
%% MAINMATTER
%%%%%%%%%%%%%%%%%%%%%%%%%%%%%%%%%%%%%%%%%%%%

\section{Introduction}

Simbol--X will be very important for the understanding of the physics
of jets by observing low power, lineless and TeV emitting BL Lacs and 
high power, Flat Spectrum Radio Quasars (FSRQs) with relevant broad 
emission lines.
Both subclasses can emit most of their luminosity in the hard X--ray band.
While low power BL Lacs are well known hard X-ray emitters, we here
emphasise the properties of the (less known) most powerful blazar jets.
They can be the most rewarding Simbol--X blazar targets.

Very low and very high power blazars are at the two extremes of the
blazar sequence, since its main parameter
is the  apparent luminosity \cite{fossati98}.
Their spectral energy distribution (SED) has two broad humps
whose peak frequencies increase as the total luminosity
decreases. 
At the same time, the dominance of the high energy hump over the low
energy one increases with luminosity.
This sequence has been interpreted in terms of radiative cooling 
becoming more important with total luminosity, thus allowing the
presence of high energy electrons only in low luminosity sources
\cite{gg98}.
In \cite {gg08} we proposed a new version of the blazar sequence, 
linking the jet power to the accretion rate
of the associated disk, and pointing out the importance of the
mass of the black hole, governing the distance scale $R_{\rm diss}$
where most of the jet dissipation occurs.
This is a crucial quantity, since it determines the value of the magnetic field
in the emitting region, and controls, in high power blazars, the amount of
radiation produced externally to the jet, whose photons  
are the seeds for the ``external" inverse Compton (EC) process. 
Thus the shape of the SED and the ratio between the inverse
Compton to the synchrotron luminosity (called ``Compton dominance")
depends on $R_{\rm diss}$, as investigated in \cite{gg09}.
As a source of seed photons, we included the radiation from
the accretion disk and its X--ray corona, the broad line region (BLR),
an infrared emitting torus and the cosmic microwave background.

\section{Low power BL Lacs}

\begin{figure}
\includegraphics[height=.35\textheight]{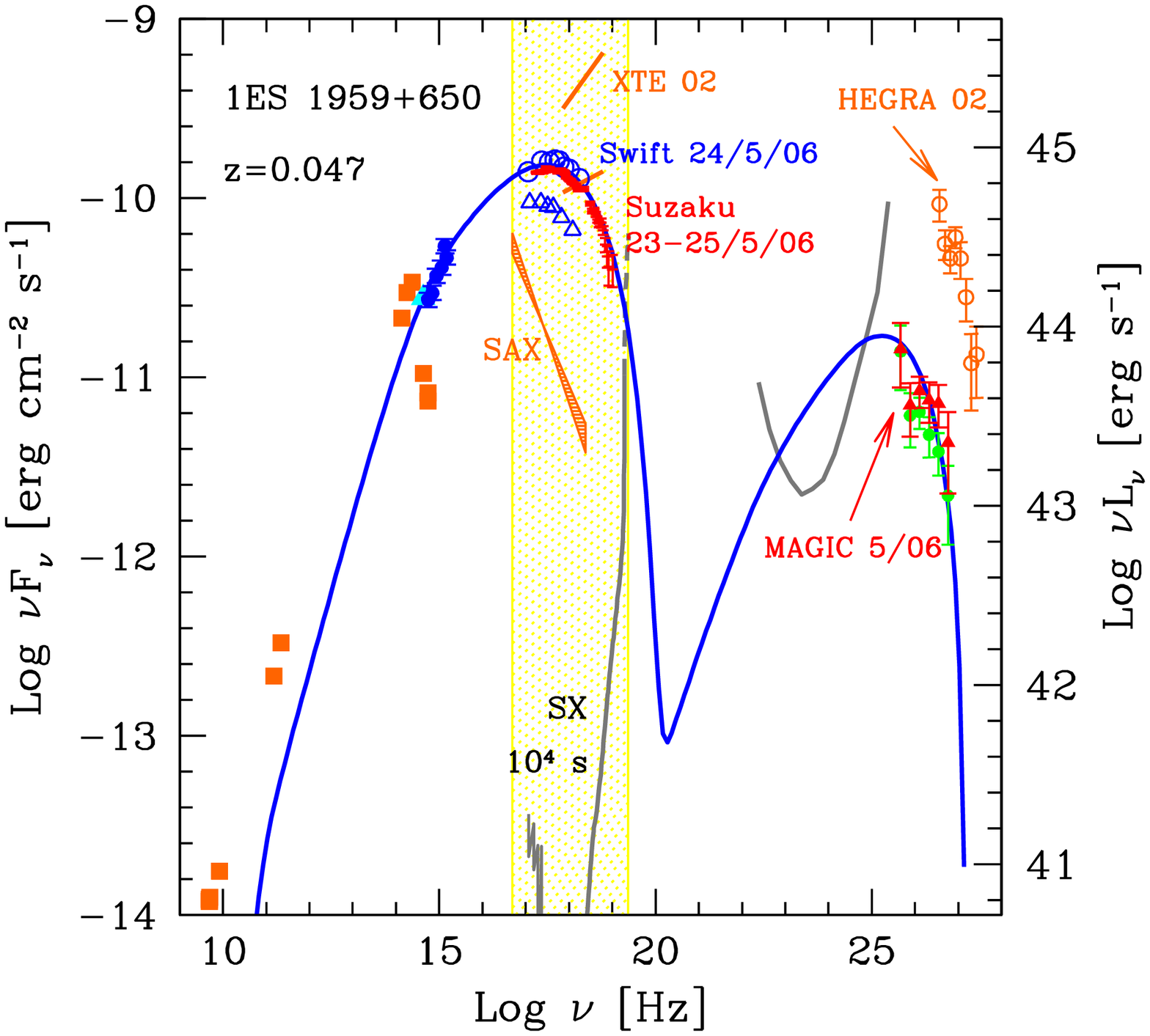}
\includegraphics[height=.35\textheight]{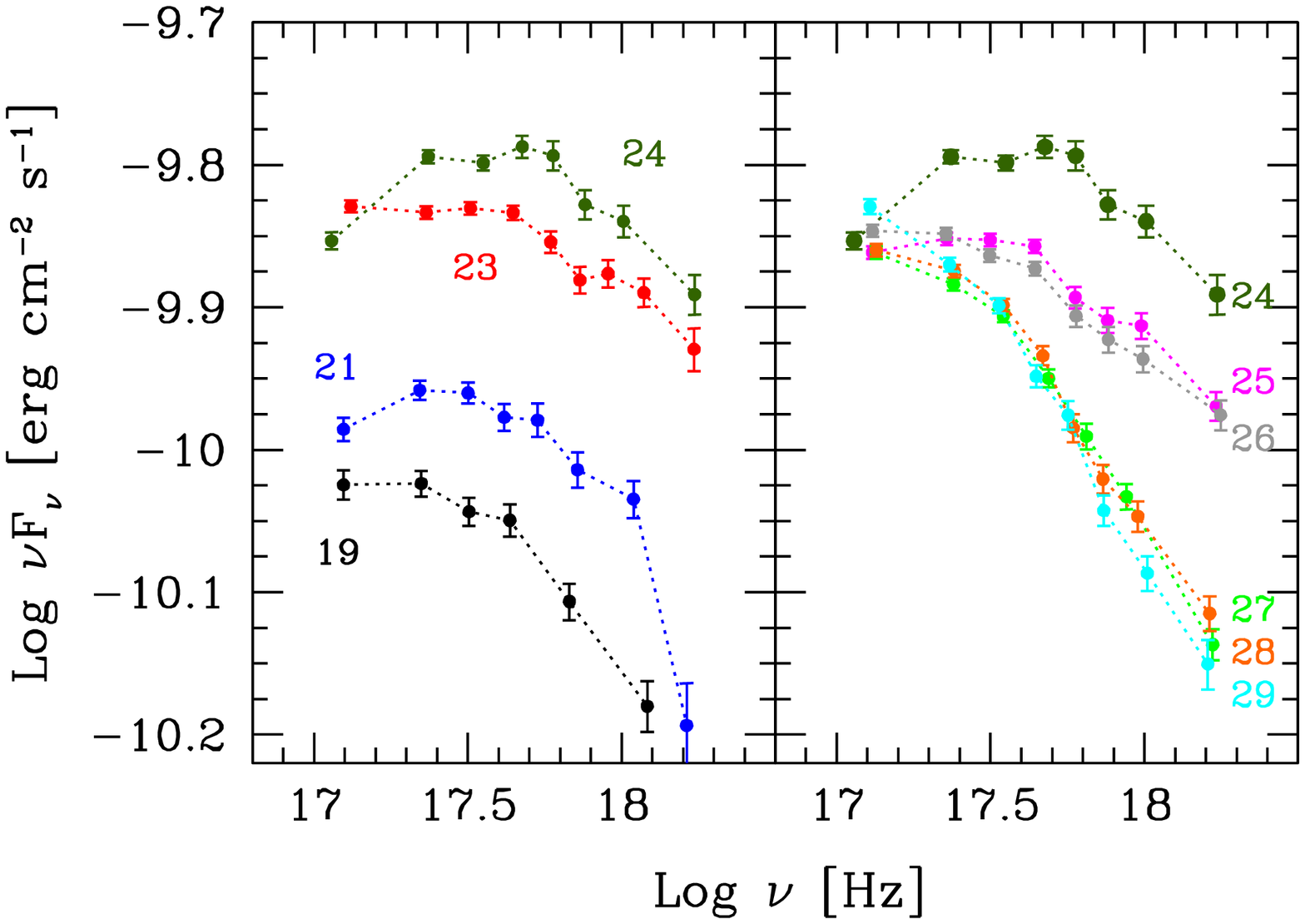}
\caption{{\it Left:} the SED of 1ES 1959+650 (data from \cite{gt08} 
and references therein).
We show also the sensitivity curve of Simbol--X for $10^4$ s exposure. 
{\it Right:} The X--ray spectra at different days during the campaign 
of May 19-29, 2006, from \cite{gt08}. On the left panel the flux increases; while on the right
panel it decreases.
On this week--timescale, the variability has a well defined trend, and it is not random.
}
\label{f1}
\end{figure}

Low power jets, and their progenitor FR I radio--galaxies, are likely 
associated to radiatively inefficient accretion disks, accreting
at a small fraction (i.e. $\sim 10^{-3}$) of the Eddington rate.
Thus the ionising radiation is small, and the BLR, even if it can be present
(as in BL Lac itself), should be located very close to the black hole.
Dissipation distances $R_{\rm diss}$ of order of a few hundreds of \sc\ radii
imply that most of the jet luminosity is emitted in a ``photon clean"
environment, with no (or weak) external seed photons (apart from the possible
presence of an emitting slow ``layer", see below).
Consequently, the high energy emission is due to the self--Compton
mechanism (SSC). 
Fig. \ref{f1} shows one example, 1ES 1959+650 
(data in \cite{gt08}).
The synchrotron flux peaks at a few keV, ensuring the presence of
high energy electrons, producing, by their SSC emission, a peak 
at hundreds of GeV. 
There are other, even more extreme BL Lacs, like Mkn 501, 1ES 1101--232
\cite{pian,aharonian,wolter},  
with synchrotron peaks above 100 keV.
Low power BL Lacs are therefore the best TeV candidates.
Fig. \ref{f1} shows also the sensitivity curve of Simbol-X 
for an exposure time of $10^4$ s, to illustrate that even
with a relatively short exposure it is
possible to study the spectrum and the variability
of these kind of sources, up to 100 keV.
Among the most interesting topics to be studied there are:
i) to find possible lags between different X--ray frequencies,
including the unexplored 10--100 keV range.
This gives information on the cooling/acceleration process
(i.e. cooling electrons should produce lags between high and low
frequency fluxes, while shock accelerated electrons should correspond
to the opposite behaviour, see e.g. \cite{kirk}.
ii) We can study trends in the X--ray spectral variability. 
This can shed light on the way the dissipation occurs: if the
variability has a random character, then it can be produced by 
different emitting region varying independently. 
This is what the ``internal shock model" predicts (\cite{spada,guetta}.
If there are well defined trends, then a ``standing shock model"
is favoured (see, e.g. the discussion in \cite{gt08} and Fig. \ref{f1}).
iii) Electrons emitting at 10--100 keV by synchrotron emit around or above the 
peak of the SSC one. Coordinated campaigns can tell if both fluxes are coming from
the same electrons. 
iv) If so, the detailed knowledge of the synchrotron spectrum helps to
model the high energy one, and therefore to put constraints on the
amount of $\gamma$--$\gamma \to e^{\pm}$ absorption suffered by 
high energy photons from the IR cosmic background.

\section{High power blazars}

We now turn to the other extreme of the blazar sequence, 
stretching it to very large values of the jet power.
It is sensible to measure jet powers in
units of the Eddington luminosity of the associated accretion disk 
(see e.g. \cite{gg08}). ``Large" therefore means close to Eddington.

\subsection{Powerful blazars with different black hole masses}

\begin{figure}
\includegraphics[height=.35\textheight]{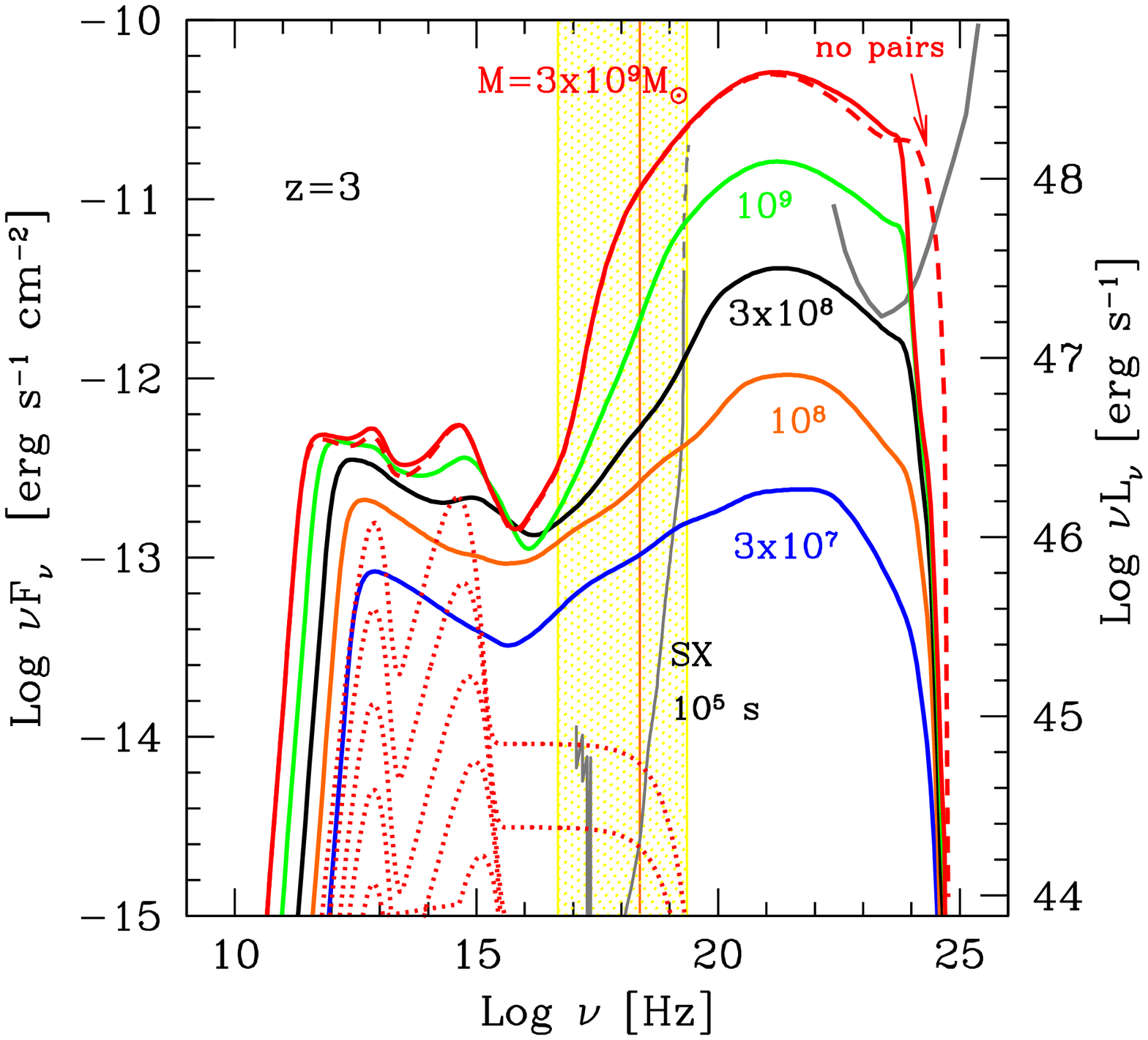}
\includegraphics[height=.35\textheight]{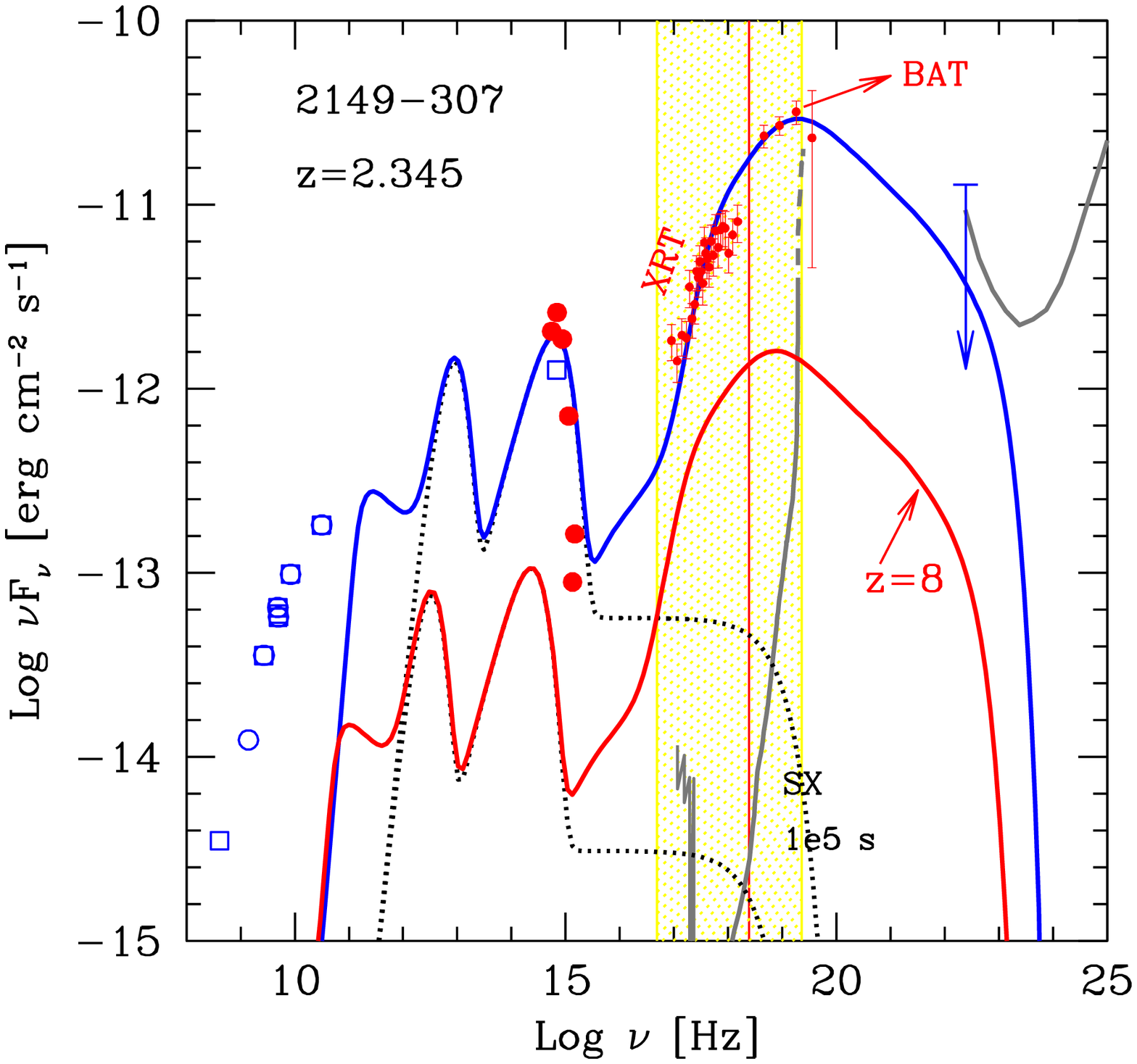}
 \caption{
{\it Left:} predicted SED as a function of the black hole mass.
We assume that the dissipation occurs at the same distance
(in units of the \sc\ radius $R_{\rm S}$) and that the injected luminosity
is the same (in units of the Eddington one). 
Also the disk luminosities (shown by the red dotted lines together with the 
IR emission from the torus) are the same in units of the Eddington one.
All sources are assumed to be at $z=3$.
{\it Right:} the SED of the blazar PKS 2149--307 (data in \cite{sambruna07})
together with a fitting model (blue upper line) and the same model redshifted
to $z=8$ (red lower line).
The high energy hump of these very powerful blazars peak in the 
100 keV--1 MeV energy range, and are therefore good targets for Simbol--X,
that can characterise the high energy peak even if the source were at $z=8$.
In both panels the grey line is the Simbol--X sensitivity for an exposure of $10^5$ s.
}
\label{f2}
\end{figure}

For illustration, the left panel of Fig. \ref{f2} shows 
the predicted SED produced by FSRQs
with different black hole masses (from $3\times 10^7$ to $3\times 10^9 M_\odot$)
scaling $R_{\rm diss}$ and the jet power with
the \sc\ radius and the Eddington luminosity, respectively.
The Poynting flux $L_{\rm B}$ carried by the jet is a fraction of its
total power $L_{\rm j}$, assumed to be constant all along the jet, so 
$L_{\rm B} \propto R_{\rm diss}^2 \Gamma^2 U_{\rm B}\propto L_{\rm j}\propto M$
which follows from $L_{\rm j}/L_{\rm Edd}=$const.
We then have  
$ U_{\rm B} \propto  L_{\rm j}/ R^2_{\rm diss}  \propto M^{-1} $
since we assume $R_{\rm diss}/R_{\rm S}=$const.
For all our cases, the dissipation occurs within the BLR, which yields a constant
radiation energy density $U^\prime_{\rm BLR}$ 
(we assume $R_{\rm BLR}\propto L_{\rm disk}^{1/2}$ and the same $\Gamma$ for all sources).
As a consequence, the ratio 
$L_{\rm EC}/L_{\rm syn} \sim U^\prime_{\rm BLR}/U_{\rm B} \propto M$.
This is the reason of the increasing dominance of the inverse Compton 
emission increasing the black hole mass.
This implies that blazars with large black hole masses
should preferentially be more Compton dominated, and therefore 
more easily detected by the Fermi satellite. 
In fact, in Fig. \ref{f1} one can see the 5$\sigma$ sensitivity
of Fermi for 1 year of operation (grey line), suggesting that, 
at high redshifts,
the detected blazars will preferentially have large black hole masses.

Also the importance of the EC relative to the SSC emission increases 
with the black hole mass, hardening the X--ray spectral shape.
For the SED with $M=3\times 10^9M_\odot$ we show the effects
of neglecting the $\gamma$--$\gamma$ absorption and the
consequent reprocessing (dashed line). The effect is modest, since
the primary spectrum breaks at $\sim 10$ GeV energies due to 
the decreasing, with energy, Klein--Nishina scattering cross section.
This figure shows that there are many high power jets with
intermediate black hole masses that are undetectable by Fermi, 
but are well visible by Simbol--X.
It will be hard for Simbol--X to serendipitously
discover these blazars, due to its small field
of view, but good targets for Simbol--X can come from the BAT/Swift 
and/or the INTEGRAL surveys, that will select blazars bright in the hard 
X--ray range.

\subsection{Up to $z=8$}

Interesting results concerning high redshift, powerful blazars 
have been already obtained by \cite{sambruna07}
with BAT; by \cite{bassani07} with INTEGRAL and by \cite{tavecchio07}
with Suzaku. 
One interesting blazar present in the 9--months BAT survey \cite{sambruna07}
is PKS 2149--307 at $z=2.345$.
The right panel of Fig. \ref{f2} shows that the high energy peak lies
in the 100 keV--MeV range.
This figure shows also the the best fitting model
(blue upper line) corresponding to $R_{\rm diss}=\sim 10^{18}$ cm,
(i.e. 800 $R_{\rm S}$ for the assumed black hole mass of 
$M=4\times 10^9 M_\odot$)
The BAT data have large error bars, precluding the possibility to firmly claim that
its high energy peak is within the BAT energy range (namely, at $\sim 100$ keV),
but this possibility is indeed suggested by the present data.
The Simbol--X sensitivity for an exposure of $10^5$ s is shown by the
grey line. 
The red lower line corresponds to the same model used to fit the blazar,
but now placing it at $z=8$, to demonstrate that even
at these redshifts Simbol--X has the capability to detect and study 
this blazar, and, most importantly, find its high energy peak.

The main scientific issues to be studied include:
i) we can determine the jet power in the most powerful
sources, by the knowledge of the peak of the SED, that
ii) will also depend upon $R_{\rm diss}$ 
thus giving informations on the mechanisms of jet
dissipation.
iii) The coordinated variability will tell if simple
one--zone models are viable. 
iv) High power blazars
should have a visible UV bump, allowing to directly
study the jet/disk connection and the ratio of their
respective powers.

\begin{figure}
\includegraphics[height=.35\textheight]{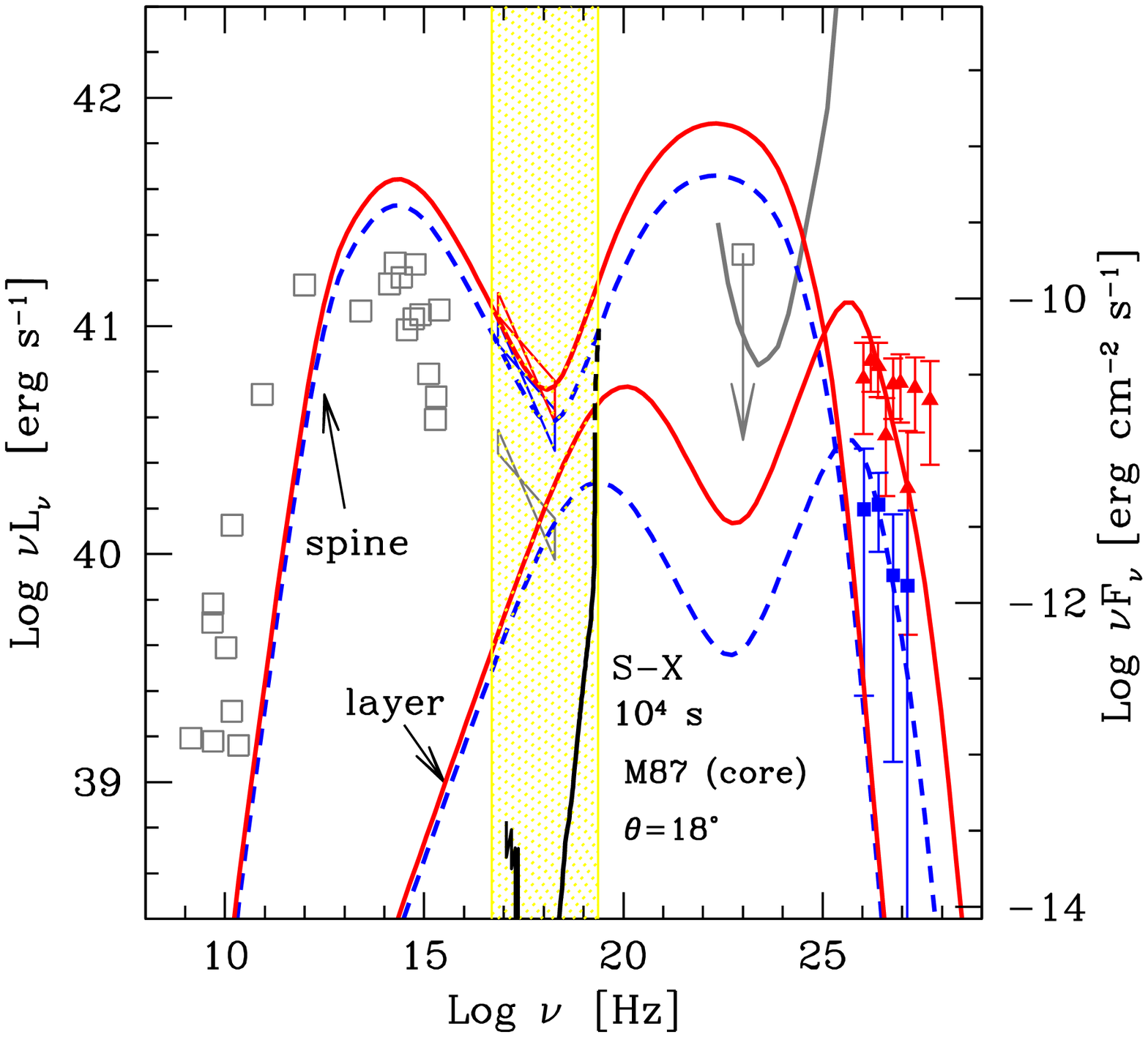}
\includegraphics[height=.35\textheight]{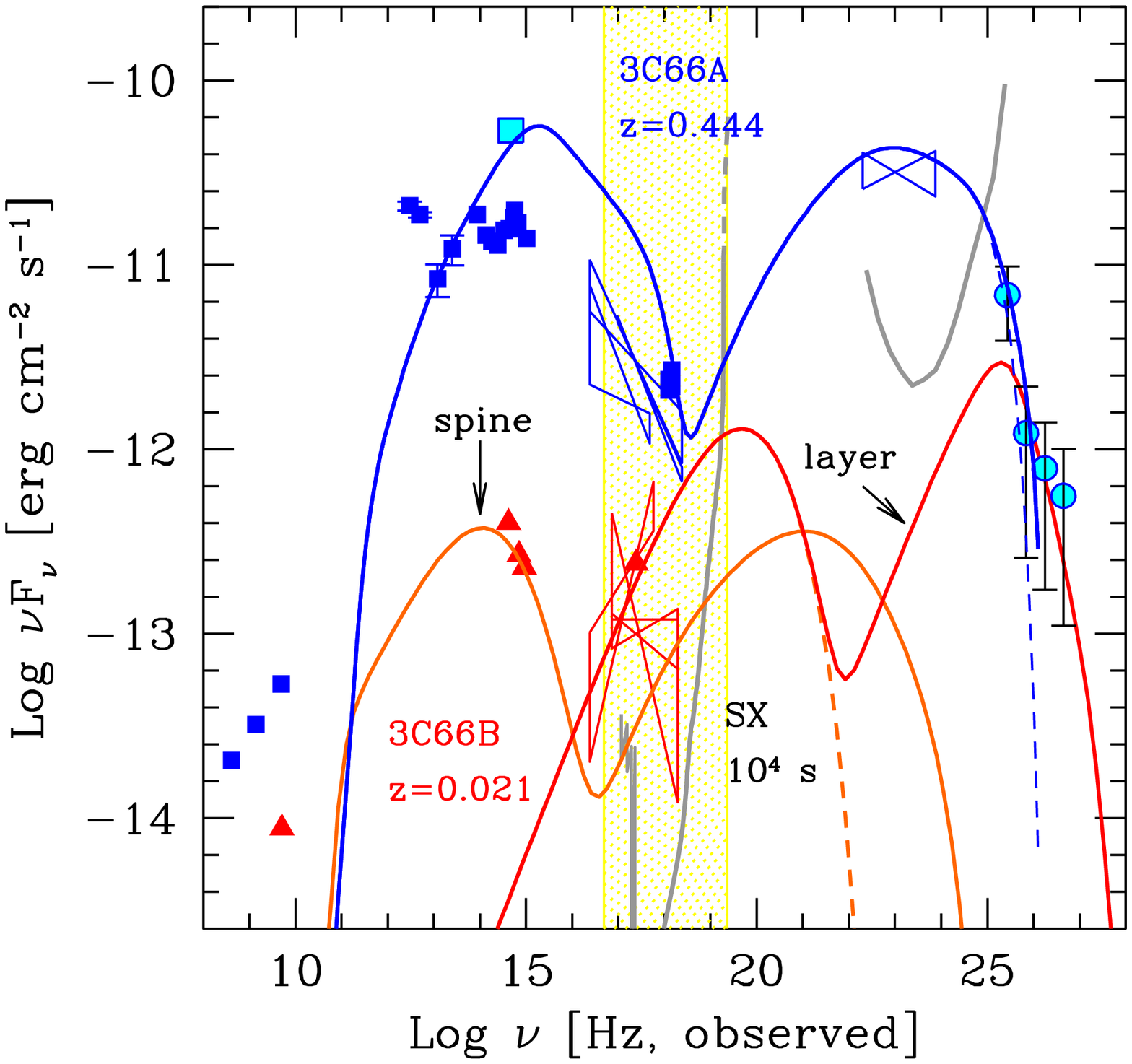}
\caption{
{\it Left:} SED of the core of M87 together
with the H.E.S.S. spectra taken in 2004 (open blue squares) and
2005 (open red triangles), from \cite{aharonian06}.  
The lines report the emission from the spine and from 
the layer for the two states. Adapted from \cite{tavecchio08}. 
{\it Right:} SED of 3C66B and of 3C66A.  
The measured VHE spectrum of MAGIC (cyan circles, \cite{aliu09}), 
is reproduced as the sum of the
emission from both sources, 3C66A being dominant below 200 GeV, and
3C66B accounting for the emission above 200 GeV. 
Adapted from \cite{tg09}.
}
\label{f3}
\end{figure}

\section{Radio--galaxies}

The typical bulk Lorentz factor $\Gamma$, in blazars, is in the 10--20 range, and
there seems to be the need, in flaring TeV blazars, to have even more
extreme $\Gamma$ around 50 to overcome spectral and variability 
difficulties \cite{konopelko, begelman}.
Therefore the typical beaming angle $\theta_{\rm v}\sim 1/\Gamma <5^\circ$.
Radio--galaxies, being observed at much larger angles, should be therefore
strongly de--beamed.
On the other hand, there is the possibility that the jet has a velocity structure,
being faster in its spine, and being surrounded by a slower layer.
If this layer is emitting, there is an interesting radiative interplay between 
the spine and the layer: both components see the emission from the other amplified
by beaming (due to the relative motion).
The extra photons are used as seeds for the inverse Compton emission, which is
then more intense for both the spine and the layer.
In \cite{gtc05} we have studied the SED resulting from assuming a fast spine--slow layer
jet structure, finding that i) radio--galaxies can be relatively strong 
high energy emitters, and ii) that when the radiative interplay between the 
layer and the spine is strong, the latter can decelerate, due to Compton drag,
helping to explain the small apparent VLBI velocity \cite{piner08}.

We have then applied this model to two radio--galaxies observed in the TeV range
(M87 and 3C 66B, \cite{tavecchio08, tg09}, whose SED are shown in Fig. \ref{f3}.
According to the spine/layer interpretation, the SED of M87 is dominated by
the spine emission at all but the TeV energies, where the layer dominates.
The predicted X--ray spectrum shows an upturn at $\sim$10 keV, that Simbol--X
can confirm (or not) with a relatively inexpensive exposure of $10^4$ s
(see the plotted sensitivity curve on the same figure).

For 3C 66 (A and B), the origin of the high energy emission is still debated \cite{aliu09}, 
since the two sources are separated by 6', close to the angular resolution of MAGIC.
If our model is correct, the synchrotron flux of 3C 66B should peak
at $\sim$100 keV, and this is can again be confirmed by Simbol--X with a 
relatively short exposure of $10^4$ s. 
In this case the layer dominates the emission above the UV band.

% \section{Conclusions}

\begin{theacknowledgments}
I thank the partial funding by a 2007 PRIN--INAF grant and useful discussion with Fabrizio
Tavecchio.
\end{theacknowledgments}

\end{document}